# Logic and Theory of Representation

Arnaud Plagnol

**preliminary draft - not to be transmitted without consent of the author**

**Abstract** Underlying the theory of inferences, a primary task of logic is language analysis. Such a task can be understood as depending on a general theory of representation, taking as a starting point the idea that some entities (« representations ») can *present* some entites (« contents »). We outline a theory of representation accounting for the capacity of representational systems to access universes that extend beyond an immediate presence. We define three logical properties that any adequate representational system should have: completeness, faithfulness, coherence. We show that logical laws are laws of representation. Finally, it appears that logic can be considered as the abstract theory of representation.

**Mathematics Subject Classification** Primary 03A05 – Secondary 03B05

## 1 Introduction

The task of logic today is mainly conceived as the theory of reasoning, especially the theory of valid inferences. However, a task of language  analysis is intertwined to this first task. Indeed, without an adequate way to express the thoughts or/and the states of the world(s), any theory of reasoning would be put in risk. The modalities by which some phrases represent a content (or not) are so crucial for the validity of an inference that the formal disposition of the premises and conclusion determine this

---

I am grateful to Tony Wards for proofing the English manuscript.

validity. The works of pionneers, like Frege and Russell, have forcefully shown the necessity of this langage analysis, so this second task is in fact prior to the first, leading to the development of formal languages, typically constructed by defining a syntax before to bring in a semantics.

At the heart of this second task of logic is the idea of *representation*: a language — be it natural or formal, linguistic or mental — can be conceived as a representational system. What is a representation ? The term « representation » is so overloaded with a plurality of meanings that any re-use can be confusing. In this paper, asking the reader to completely forget any prior use of « representation », we adopt a simple and general approach: the basic idea underlying re*present*ation is that some entities (« representations ») can *present* some entities (« contents »). (For example, a picture can present a lion, a proposition can present a thought or a state of the world…) Of course it is still an inchoate idea, but we can explore the directions that it suggests and progressively refine it.

The aim of this paper is to follow the logical thread suggested by this starting point in order to show that logic can be usefully considered as the abstract theory of representation. First, we will briefly show how a general theory can be constructed from this starting point (§ 2); secondly, we will specify some logical properties of a representational system (§ 3); in the end, we will show how the theory of inferences itself can be defined within such a general theory of representation (§ 4). To give an advance taste of this final point, let us say for now that an inference can be conceived as a constrained bond between the presence of one or several entitie(s) (presented in the premises) and the presence of an entity (presented in the conclusion) — from the former, we get the later —, this contrained bond depending on an abstract law of representation.

## 2. A (brief) theory of representation

How an entity E (human, machine…) can accesss to a universe U? Except if U is trivial for E, U cannot be wholly given in an immediate presence to E, so E must have a representational system to presentify U by fragments, but also to virtually unify these fragments.

### 2.1 Representational systems

Definitions 1-4. A set of entities S is a *representational system* for a universe U if U can be reconstituted from S, at least virtually, according to a *representational function* that associate some elements of U (*contents*) to some elements of S (*representations*).[1,2]

Notation 1. In the sequel of the § 2, "S" will denote a representational system, whatever it is.

Comment 1. The representational function can be implicit, as is the case for any natural representational system (e.g., the perceptive system of a living being).

Now we are confronted by the main logical puzzle about representation: the *presence's puzzle*: how a representation can *presentify* a content? Indeed, if an entity can only present itself, any content would be trivial and any representation would be without interest; but, to presentify another entity than itself would realize a logical wonder (to be different from itself).

Logical puzzle, logical answer! An entity E cannot presentify another entity than E. A representational system must therefore include two types of representational components, the ones that present directly themselves as contents, the others that have a content only by the mediation of the frst.

Defs. 5-6. A representational component is *analogical* if and only if its content is direct (i.e., immediately present, by « immersive identity » or by « acquaintance »). A representational component is *symbolic* if and only if its content isn't direct.

Principle of analogical mediation. The contents of symbolic components of representation, to be accessed, need to be analogically presented (i.e., directly displayed).

Comment 2. There is a well-known duality between two types of empirical representations: (1) « analogical » (or « iconic » or « depicting »…) representations, like images, that are similar in some way to their content, (2) « symbolic » (or « numerical », « digital »...) representations, like linguistic representations, that may have no similarity with their content. Within our conceptual framework, an analogical component of representation directly displays its content, which is nothing than itself, and a symbolic component of representation needs an analogical mediation to present a content. Any empirical representation (percept, image, map, name, proposition..) is a combination of analogical and symbolic components as defined above (defs 5-6.).

---

[1] In our definitions [defs.], the italized words are the defined items.

[2] A representational function is a relation that isn't necessarily functional in the mathematical sense.

A trivial empirical observation justifies the definitions 5-6. What analogically represents a content E within an empirical representation R (e.g., an image of E) is what is present of E in R, that is, not only what is similar to E, but what is identical to E. For instance, a photograph P of Alice analogically represents Alice insofar as P contains a figure identical to Alice's figure, and what P hasn't of Alice — e.g., the fineness of the texture of her flesh — isn't analogically represented in P.

For linguistic and mental representational systems, the analogical mediation for the symbolic components of representations has been confirmed by the rich experimental corpus of *grounded/embodied cognition* [2-4, 7]. In particular, some proposed sets of tools like *mental models* [6], *conceptual metaphors* [7], *perceptuals symbols* [2], *proxytypes* [9], show how analogical components of representation allow an access to any entity, even to a highly abstract entity (like a mathematical entity). In fact, taking into account some logical arguments, like the presence puzzle or the symbol grounding problem [5], analogical mediation is a logical necessity [8].

The presentation of symbolic components depends also on analogic mediation, as is well-known in logic.

Def. 7. A *syntactic representational component* is an analogical component that presents some symbols as its content.

Comment 3. Even processes that seem purely symbolic, like formal arguments, lie on analogical components of representation that are the only access to any content. Indeed, in a formal demonstration, statements work in a pure syntactic way. The purpose of such a demonstration is to secure the validity of a conclusion thanks to a chain of immediate obvious steps, so the display of the semantic content of the statements is put aside, but the demonstration is no less *presented*, the analogical displayed components being in this case some syntactic entities. Thus, the mapping of a schema of inference is nothing else than the direct grasping of an identity of form — e.g., one matches "((if it rains then the ground is wet), it rains) therefore the ground is wet" with the premises and conclusion of a *modus ponen*s schema.

The analogical capacity of a representational system (i.e., its capacity for an immediate display) has usually some strict limitations.

Defs 8-10. The *window of presence* of S is the maximal analogical capacity of S, i.e. its maximal capacity of immediate presence. An *elementary analogical fragment* [EAF] for S is an element of S that fullfills the window of presence if it is presentified. The *analogical basis* of S is the set of its EAFs.

Comment 4. A window of presence may have at least three types of limitations: (1) bounded size, (2) finite power of resolution, (3) limited number of dimensions. For example, the representational power of a television screen, a browser window of a website, or a working memory's "visuo-spatial" component of a mental system [1], is constrained by these three types of limitations.

Def. 11. The *format* of a representational system is defined by the geometrical properties of its window of presence.

Example 1. The format of the common human spatial experience is euclidian and tridimensional.

## 2.2 Symbolic systems and analogical extensions

To reconstitute some extended universes, a representational system needs to « chain » EAFs within its window of presence. Moreover the EAFs that are not displayed at a given instant in the window of presence need to be coded and stored in a *memory*. The function of symbolic representations is to ensure both the chaining and the storing of the EAFs.

Defs. 12-14. The *symbolic system* (or *web*) of a representational system S is the system of entities (*symbolic units*) that is used (1) to code the elements of the analogical basis, (2) store these elements (what constitutes a memory), (3) chain these elements in the window of presence. The *semantic function* of S is the function that attributes a content (i.e., a fragment of the represented universe) to some elements of the symbolic system.[3]

---

[3] A semantic function is a relation that isn't necessarily functional in the mathematical sense.

<u>Comment 5</u>. The content of a symbolic unit, being attributed by a semantic function, isn't direct, in keeping with the definition 6. The semantic function defines the representational function for the symbolic elements. Note also that a semantic function can assign to a symbolic unit a part or the whole of its direct content (e.g., the system of quotation marks in a natural written language).

Thus, analogic mediation allows for the presence of contents, and symbolic systems allow for the the the building of extended universes from analogical bases.

<u>Def. 15</u>. An *analogical extension* for a representational system S is a (set of) fragment(s) of universe that is reconstituted from some EAFs and some fragments of the web of S.

If an analogical extension E is not trivial, E cannnot be wholly immediately presented in the window of presence, but E can be virtually available by « navigation » in E, that is by chaining some successive EAFs in the window of presence.

<u>Example 2</u>. Assume the mental system of Tom, an homesick Londoner that lives in Paris, but often mentally walks into London. Each mental image about London that Tom can presentify from his memory is an EAF for his mental system; the analogically unified set of these mental images builds a virtual model of London within which Tom can navigate, even though Tom is unable to wholly display it at a given instant; this virtual model is an analogical extension for the mental system of Tom.

Two symmetric operations are necessary to perform the conversions between analogical displays and symbolic coding within a representational system.

<u>Def. 16</u>. An operation of *abstraction* is the formation of a symbolic unit from EAFs.

<u>Example 3</u>. A mental symbol *CAT* can be formed from percepts and images of cats.

Def. 17. An operation of *projection* is the display of an EAF in the window of presence from a (set of) symbolic fragment(s).

Example 4. A mental image of a prototypical cat can be displayed in the window of presence of a mental system from the activation of mental symbols about cats.

Def. 18. A *focusing* is the projection of an EAF from a symbol syntactically included within an EAF.

Example 5. The effect of clicking on a hypertext link on a web page is a focusing.

Remark 1. The power of abstraction determines the capacity of representation of a representational system. Several degrees can be defined:

- Degree 1 (*Naming*). Naming is the raw grasp of a recurrence with the attribution of a symbolic unit to this recurrence.[4] For instance, « rrrrhum ! » is uttered before a human situation by a (hungry) lion (assuming that « rrrrhum ! » is the equivalent of the english word « human » in a leonine language).
- Degree 2. (*Predication*). The operation of predication is an abstraction that gives rises to a proposition. With a predication, the operation of naming (degree 1) is explicited: a predication exhibits an entity (or several entities) that bears the recurrence named by the predicative (or relational) term. Indeed, in our framework a proposition is the product of an analysis of a situation in the window of presence.[5] For instance, « This is human. » is uttered before a percept of a human subject. (The operation of predication seems to be properly human – as it were, human is the propositionalizing animal.)
- Degree 3 (*Variabilization*). In our framework, a variable is a tag of abstraction: variabilization is the representation of the operation of abstraction itself. For instance, « a human » (« this, as it is human ») is uttered before a percept of a human subject, meaning that the thing before the eyes is taken as human abstracting from the specific human it is. The implicit tag of abstraction linked to « a human » becomes apparent with the notation « an x such human(x) ». From variabilization, *generalization* (« any human ») and *universalization* (« all the humans ») can be defined and used in laws.

---

[4] A recurrence is something that (potentially) recurs in a series of situations.

[5] Far from being a composition, as it is often asserted, a proposition is fundamentally a decomposition.

## 2.3 From situations to representational space

Let's bring in now some useful definitions to specify the concept of analogical extension. (We won't develop a rigorous theory, limiting ourself to give an intuitive idea of a few fruitful notions.)

Defs. 19-20. A *situation* is an EAF considered *qua* content (i.e., as a fragment of the represented universe). A *unifying representation* is an analogical extension « in one piece », so that its content is reconstituted from several situations that are chained by *one* fragment of the web.

Example 6. Consider the mental system of the Londoner Tom (see Example 2). In principle, Tom's virtual model of London is a unifying representation.

Defs. 21-24. An *unfold* is an analogical unifying representation, so that it virtually co-displays several situations whose union is a « plain » fragment of the represented universe. A fragment of the represented universe U that can be represented in one unfold is *simple* for U. A *s-unfold* is an unfold that syntaxically contains some symbols. A *w-unfold* is a wholly displayed unfold where all the symbols has been eliminated by iterative projections.

Defs. 25-28. A s*ymbolic structure* is a web fragment that allows some situations to be chained, so that it yields a means to navigate within a unifying representation. A *link* is a piece of symbolic structure that chains two situations consecutively. If a link results in an unfold that analogically unifies two situations, it is an *A-link*. If a link isn't an A-link, it is an *artificial link*. (An artificial link isn't grounded on an effective content.)

Def. 29. An *object* is a unifying representation organized around a specific symbolic unit (i.e., a coding unit that allows the connection of different representational fragments building the unifying representation).

Example 7. A mental object is constituted by a symbolic unit that unifies some series of (fragments of) situations that are given during the lifetime within percepts, images or mental models. For instance, the analogical fragments about London of the Londoner Tom (see Example 2) are structured around his mental name *London*.

Def. 30. A *path* from a situation σ to a situation σ' is a series (σ,..., σ') of situations that can be successively displayed in the window of presence, with the proviso that two consecutive situations of the series are connected by a link. (Thus, a path is a special kind of unifying representation.)

Example 8. During a daydream about London, Tom (see Example 2) can clear a mental path from the Tower to Piccadilly Circus thanks to a series of images that are successively projected in the visuo-spatial component of his working memory. (If Tom doesn't know how to overcome such or such a passage, Tom needs to connect two situations by an artificial link.)

Def. 31. The *union* of several unifying representations R, R', R''… is defined by identifying the situations and the links that are common between some of these unifying representations.

Comment 6. If an analogical fragment is common to R and R', the union of R and R' is itself a unifying representation; if no analogical fragment is common to R and R', their union consists of two separate fragments.

Def. 32. The *representational space* is the union of all unifying representations (or the union of all analogical extensions), that is, the universe effectively represented by S.

Remark 2. Many unifying representations can be constructed from the same basis of situations — just as many towns can be constructed from the same stock of Lego$^R$ elements — so that some fragments of the representational space can be contradictory.

## 3. Some logical properties of a representational system

We have outlined a theory of representational systems relative to the restitution of universes. What can be the role of logic for such a theory? Of course representational systems need to satisfy some logical constraints. To specify the ability of a representational system to ensure its function — that is, the restitution *ad integrum* and exclusive of a given universe —, we need to define three properties: completeness, faithfulness, coherence.

Notation 2. In the sequel of the § 3, "S" will denote a representational system (whatever it is) defined by a representational function R, "U" will denote the universe aimed to be represented by S, and "Σ" will denote the symbolic system of S.

## 3.1 Completeness

Completeness characterizes the capacity of a representational system to provide the whole universe for which it is made.

Defs. 33-34. S is *complete* if S is able to wholly provide any fragment of U, that is any fragment of U can be fully given within an analogical extension for S. S is *symbolically complete* (*s-complete*) if Σ is able to code any fragment of U.

Example 9. In analytic geometry, figures are reduced to equations that are supposed to be s-complete. (For instance, "$(x^2+y^2) = 1$" is supposed to wholly code a unit circle.)

Example 10. Assume the mental system of the Londoner Tom (see Example 2). If the mental system of Tom is s-complete relative to London, and under the condition of a sufficient capacity of projection, Tom can navigate anywhere in London from his sole memory.

Comment 7. Completeness can be easy to realize when U is whollly fixed and determined, for instance when it is a website (a website is s-complete by definition), or when U is the universe of a robot that stacks cubes according to a small number of factors. When U is not artificially defined, it is generally a utopian endeavour to search for a complete system, but completeness can be obtained for some partial fragments of U. When representational facts belong to U, logical reasons can prevent completeness. (Representation of representation may give rise to some paradoxes.)

## 3.2 Faithfulness

Faithfulness characterizes the capacity of a representational system S to restitute the expected universe, that is the universe U for which S has been made, and no other universe.

Defs. 35-36. S is *faithful* if any content C that is reconstituted by S from a representational fragment F is a fragment of U satisfying C= R(F). Otherwise S is *unfaithful*, that is the universe reconstituted by S is in fact a universe U' different from U.

Example 11. Assume the visual system V of the human Tom. According to its biological function, V is expected to reconstitue the visible world that can be accessed within the human format of presence, what defines a representational function. This system is faithful if every visual percept gives the expected content according to this function. If Tom walks in London, looks at the Tower and sees the Great Pyramid of Egypt, V is unfaithful.

A representational system can be unfaithful due to an analogical basis that contains erroneous data, and/or due to an inadequate symbolic sytem. Inadequacies of the symbolic system can occur at every level of abstraction (see Remark 1):
   - the entities of the universe may be ill named by the symbolic system (e.g., if two entities receive the same name);
   - the predicative/relational links between entities may be wrong;
   - the representation of domains of abstraction may give rise to errors (e.g., errors relatively to the validity domain of a law).

**3.3 Coherence**

Coherence characterizes the capacity of a representational fragment to be integrated wihin an unified analogical display.

Defs. 37-39. Let F be a fragment of a representational system S. The *explicitation* of F is the elimination of symbols from F by iterative projections. F is *coherent* if and only if its explicitation gives rise to an analogical unified display, except for the frontiers between different worlds. F is *incoherent* if F isn't coherent.

Comment 8. A fragment of representation can be incoherent without the symbolic system being able to express it.

Remark 3. If U is real, U is necessarily displayed (perhaps with some frontiers between the worlds it contains) and any incoherence comes from the representational system.

Defs. 40-41. A fragment F of a representational system S is *intrinsically incoherent* if F is incoherent because of the constitutive rules of S, otherwise F is *extrinsically incoherent*.

Comment 9. It could seem strange to make a system with representational fragments that are intrinsically incoherent, but it may be very difficult to exclude *a priori* such elements. Indeed, a typical symbolic system is built from simple syntactic rules, so it contains many symbolic fragments without content (e.g., false propositions), some of these being incoherent by the rules of the system (e.g., contradictory propositions). Some symbolic fragments can also be intrinsically incoherent because their explicitation gives rise to a vicious circle or to a regression *ad infinitum* (e.g., « I'm saying the truth » or « I'am lying » in English).

Any extrinsic incoherence comes from an unfaithfulness: erroneous data within the analogical basis, wrongness in the coding of the entities or of their setting, errors in the domain of validity of a law by a too hasty induction… Unfaithfulnesses gives rise to blockages during a process of explicitation towards a unified display, because of irreducible incompatibilites between fragments.

Example 12. Let C be a red notebook. Assume that C is coded as red in Jane's mental system, because she has already seen the notebook, and assume that C is also coded as blue in this system, because of Jane's hallucination. Some projections of C within Jane's window of presence cannot be analogically unified.

Remark 4. Any unfaifhfulness may give rise to an incoherence. (If a red note book C is registered in the system as blue, the visual discovery of the redness of C can give rise to an incoherence.)

# 4. Inferences and theory of representation

Representational systems need to satisfy some logical constraints, as some of these are set by the properties of completeness, faithfulness and coherence. But logic as the theory of valid inferences can itself be considered as included in the abstract theory of representation. To prove this point, we proceed in three steps: first, we specifiy the functions of the traditional propositional connectives as mechanisms of analogical extension; secondly, we specify a concept of general implication associated to the expression of laws; third, we show that logical laws are abstract laws of representation.

## 4.1 Elementary mechanisms of analogical extensions

Our theory of representation allows for a unified understanding of the functions of the traditional propositional connectives:

- *Conjunction*. In our framework, the conjunctive operator (« and ») means the co-presence *de jure* of two facts even when the two facts cannot appear together in the window of presence. Thus, with this very simple mechanism, it is possible to overcome the limits of an EAF.

- *Disjunction*. In our framework, the disjunctive operator (« or ») means the co-display of several alternatives relative to the values of a predicate (or relation) for a same substrate. (Each alternative corresponds to a different possible world, so the co-display is limited by the « frontiers » between the worlds.)

Comment 10. Disjunction is necessary when some alternatives to the world of reference need to be considered for an analogical extension of a set of EAFs (or a set of unifying representations). In particular, disjunction is necessary to overcome epistemic limits. For example, the real world isn't wholly known and displayed to a human scientist, because of the limits of the human window of presence, thus, before being able to obtain the right analogical extension for a given set of data, a human scientist needs to consider a set of alternatives (hypotheses) and use a disjunction.

- *Negation*. The fundamental use of negation is here understood as meaning an incompatibility (i.e., an impossibility of analogical unification) of what is negated with a situation of the world of reference.

<u>Comment 11</u>. The use of a negation is necessary when there are some alternatives to the world of reference W, which need to be coded as different from W (or from each other).

- *Implication*. In our framework, the basic use of implication isn't the material implication of the usual formal logic, but the *particular implication*. A particular implication means a constraint relative to the copresence *de jure* of the consequent with the antecedent. Thus, « A implicates B » can be paraphrased by « Given A, B » (or even better: « From A, B. ») A particular implication appears therefore as a symbolic « shortcut » of an analogical scenario leading from the antecedent to the consequent, what is an extrordinary means of analogical extension.

<u>Example 13</u>. « If it rains on the White Mountain, there is a flood at Blackstone. » can be paraphrased as « Given that it rains on the White Mountain, there is a flood at Blackstone. »  This implication provides a shortcut of the scenario that unifies the rain on the White Mountain and the flood in Blackstone. If I know this implication and I learn that it rains on the White Moutain by listening to the weather, I can display in my window of presence a flood situation in Blackstone, without knowing anything between the two spots.

<u>Remark 5</u>. If the antecedent is false, a particular implication has no content: nothing is given and it it impossible to « jump » to a consequent from nothing.

In our framework, *the connectives can be extended beyond their propositional use to bear on any type of entities, whatever there are*. Indeed, their function as defined above doesn't depend on the structure of propositions.

<u>Example 14</u>. A particular implication can be defined between two objects picked up by singular terms : « If DNA fragment number 6, suspect number 535. » (or better : « From DNA fragment number 6, suspect number 535. »)[6]

---

[6] We could have said in the Example 13: « From rain on the White Mountain, flood at Blackstone. »

## 4.2 General implications and laws

A general implication is a variabilized particular implication.

Example 15. « From rain at x, flood at x »

Comment 12. A true general implication is a law. An instanciation of a law gives a particular implication. So the truth of a particular implication can be obtained from a general implication without knowing the particular scenario that is associated to it.

Example 16. Assume the biocriminal law « If it is the DNA fragment x, the donor of x is the murderer ». For each instanciation of this law, the presence of the antecedent ensures the presence of the consequent. For instance, if I know this law and the lab obtains the DNA fragment 6 that corresponds to Danny, I am sure that Danny is the murderer, even if I know nothing about the particular scenario that analogically unifies the two facts.

Remark 6. Remembering the basic idea underlying representation (see the Introduction), we can realize that *in a particular implication, the antecedent represents the consequent*: from the presence of the antecedent we get the consequent. Thus, a law L — say for instance (Px ---> Qx) —, can be conceived as a representational system S defined by a representational function f, with the equivalences between the statements (1)-(4) below:

(1) (Pa ---> Qa) by L
(2) (given Pa, Qa) by L
(3) Pa represents Qa according to S
(4) The image of Pa by f is Qa.

## 4.3 Inferences as laws of representation

Now we have the tools to interpret an inference in our framework. For simplicity, we only consider the propositional calculus.

Def. 42. A *particular inference* is a true inferencial statement that expresses a particular conclusion from particular premises within a particular world.

Example 17. Assume a novel where the propositions « If it rains on the White Mountain then there is a flood at Blackstone » and « It rains on the White Mountain » are true. « ((If it rains on the White Mountain, there is a flood at Blackstone), it rains on the White Mountain) therefore there is a flood at Blackstone » is a particular inference for the novel world.

Assume P is a particular inference for the world w1 with the premises (A1,.., An) and the conclusion C. In our framework, P is a particular implication: « [For the world w1], given (A1, A2,…, An), [one gets] C ».

Remark 7. The particular world in play is often implicit in a particular inference.

As a particular implication, a particular inference expresses a constraint link between the antecedent and the consequent. We are going to see that this constraint link in a particular inference depends on a law of representation.

Def. 43. A *particular law of inference* is obtained by variabilizing the (often implicit) world of a particular inference.

Comment 13. For the propositional calculus, the set of possible words (i.e., the domain of the variable of worlds) is defined by the valuations that attribute truth values to elementary propositions.

A particular law of inference L expresses a particular conclusion from particular premises for any possible world/valuation: « For any w, given (A1, A2,…, An), C » (w is a variable whose values are possible worlds/valuations).

Remark 8. In the last paragraph, L can also be interpreted as « For any w, (A1, A2,…, An) represents C according to a law of representation » — see below and Remark 6.)

Def. 44. A *generalized law of inference* (or *logical truth*) is obtained by variabilizing the elementary propositions, or any substituted symbolic representations (see the last paragraph of section § 4.1), in the premises and the conclusion of a particular law of inference.

A generalized law of inference is valid for any class of statements that have the same disposition of connectives between propositions (or, more generally, between symbolic representations). Generalized laws of inference are therefore some abstract laws of representation, indifferent to the particular contents of symbolic representations and uniquely determined by the functions of connectives as mechanisms of analogical extension (§ 4.1).

## 5 Conclusion

Return to the two tasks of logic raised in the Introduction. We have considered the language analysis under the perspective of a general (i.e., abstract [7]) theory of representation (§ 2), and nothing prevents us from attributing the building of such a theory to logic as its first task — what other science could be qualified for such a task as it is apparent in § 3? Moreover the second task of logic, the theory of valid inferences is included in a general theory of representation as we have seen in the § 4. Thus, we can fruitfully consider logic as the abstract theory of representation.

## References

1. Baddeley, A. D.: Working memory. Oxford University Press, Oxford (1986)

2. Barsalou, L. W.: Perceptual symbol systems. Behav Brain Sci 22, 577-660 (1999)

3. Barsalou, L. W.: Grounded cognition. Annu Rev Psychol 59, 617-645 (2008)

4. Glenberg, A. M.: What memory is for. Behav Brain Sci 20, 1-55 (1997)

5. Harnad, S.: The symbol grounding problem. Physica D 42, 335-346 (1990)

---

[7] See the remark 1 [Degree 3]).

6. Johnson-Laird, P. N.: Mental models: towards a cognitive science of language, inference, and consciousness. Cambridge University Press, Cambridge (1983)

7. Lakoff, G., Johnson, M.: Philosophy in the flesh: the embodied mind and its challenge to Western thought. Basic Books, New York (1999)

8. Plagnol, A.: La fondation analogique des représentation. Phd Dissertation, Department of philosophy, Université Panthéon-Sorbonne, Paris (2005)

9. Prinz, J. J.: Furnishing the mind: concepts and their perceptual basis. MIT Press, Cambridge (MA) (2002)

A. Plagnol
LPN (Université Paris 8) & IHPST
e-mail: arnaud.plagnol@univ-paris8.fr